\documentstyle[aaspp4,12pt]{article}

\def\clock{\count0=\time \divide\count0 by 60
     \count1=\count0 \multiply\count1 by -60 \advance\count1 by \time
     \number\count0:\ifnum\count1<10{0\number\count1}\else\number\count1\fi}

\begin{document}
\newcommand{\rg}{\sqrt{g}}
\newcommand{\etal}{{\it et al }}
\newcommand{\Nabla}{\bigtriangledown}
\def\lesssim{\mathrel{\hbox{\rlap{\hbox{\lower4pt\hbox{$\sim$}}}\hbox{$<$}}}}
\def\gtrsim{\mathrel{\hbox{\rlap{\hbox{\lower4pt\hbox{$\sim$}}}\hbox{$>$}}}}
\let\la=\lesssim
\let\ga=\gtrsim

\title{The Universal Deceleration and Angular Diameter Distances to
Clusters of Galaxies}
\author{Ue-Li Pen}
\affil{Harvard Society of Fellows and Harvard-Smithonian Center
for Astrophysics}
\altaffiltext{1}{e-mail I: upen@cfa.harvard.edu}


\begin{abstract}

We show how the virial theorem can be applied to the hot gas in
clusters of galaxies to obtain a yardstick, which could then be used to
determine cosmological parameters.  This yardstick relies on the
assumptions of hydrostatic equilibrium and that the gas fraction is
approximately constant.  The constancy is checked empirically from a
local population of clusters.  By using the observed parameters
consisting of temperature, surface brightness and radial profile
$\beta$, one can calculate the expected core radius.  Comparing it to
the observed angular size, one can in principle calibrate the
cosmological deceleration parameter $q_0$.  We test this method on a
small sample of 6 clusters, and show its promise and accuracy.  The
preliminary implications would be to suggest $q_0 \approx 0.85\pm 0.29$
with $1-\sigma$ statistical error bars, with several systematic
uncertainties remaining.  Taken at face value, this would argue against
a cosmological constant.  The method is robust to errors in the
measurement of the core radius as long as the product of the central
density and the core radius squared $\rho_0 r_c^2$ are well
determined.  New lensing and X-ray data can dramatically improve on the
statistics.

\end{abstract}

\section{Introduction}

According to the standard model of the hot big bang
(\cite{peebles93}), the universe began with small primordial
fluctuations which grew through gravitational instability.  The model
invokes the existence of gas and dark matter, which on scales larger
than $10^6$ solar masses obey the same dynamical equations in linear
theory.  After the perturbations grow non-linear at a redshift
$z\approx 10$ when the universe was a few percent of its current age,
our understanding of structure formation becomes incomplete.  Stars,
quasars and galaxies form from the collapsing and cooling gas, while
the dark matter is non-linearly stable and remains in halos around the
galaxies.  At some point clusters of galaxies formed, containing more
mass in hot gas than in stars or galaxies.  Because most are very hot
($T\approx 10^8$K), that gas is unable to cool and contract efficiently,
and would remain in hydrostatic equilibrium balanced by pressure and
gravity in the same way as the dark matter.  Detailed analyses
(\cite{white91}) show that we know of no mechanism to keep the gas out
of the potential well, nor to collapse it significantly relative to
the dark matter, and we would thus think that the gas would constitute
the same fraction of the mass in clusters as they did globally, except
for the small fraction which condensed to form stars.  This is
confirmed in numerical simulations (\cite{pen96,nav94,cen96}).  We
parametrize the Hubble constant as $50 h_{50}$km/sec/Mpc, where
$h_{50}$ is believed to lie between 1-1.6.  The first attempt to use
this hypothesis was by Henry and Tucker (1979), who derived a gas
fraction of 18\% and suggested its use for cosmological purposes.  In
this letter we use the better data collected since then to actually
estimate the deceleration parameter $q_0$.

By analyzing a local sample of clusters, we find the gas fraction
$f_g$ from an Einstein IPC sample of 5 clusters together with COMA to
be $f_g=0.178\pm 0.015 h_{50}^{-3/2}$, with an intrinsic scatter of
only 20\%.  The estimate assumes that the gas fraction is constant at
all radii.  This gas fraction is consistent with earlier estimates
(\cite{white91,whi95}).  While the constant gas fraction assumption
certainly introduces errors, we find empirically that this error must
be less than the 20\% observed in the data.  Furthermore, by
assumption this fraction does not depend on the radius at which is is
measured.  The observed radial profiles are determined from X-ray
data, which puts most weight near one core radius.  There we would
expect clusters to be well relaxed, and the gas fraction to be
representative of its global value.

In this article we use a detailed model of hydrostatic equilibrium to
calculate the core radius of clusters from their observable
properties, assuming a constant gas fraction.  By comparing it with
the angular size, we obtain the angular diameter distance to the
cluster, which provides a direct measure of the deceleration parameter
$q_0$.  Taken at face value, the small test sample suggests $q_0 =
0.85 \pm 0.29$ if we assume the sample of three distant clusters at a
redshift of $z\sim .5$ to have the same gas fraction as the local
sample.  This preliminary result is still subject to large systematic
uncertainties.  The situation is expected to improve rapidly as
current X-ray data at intermediate redshift $(z\ga 0.4)$ is collected
and reduced.  The value determined with this method is consistent with
recent determinations by Perlutter (1996a,b) using type 1a
supernovae.

We use the notation $\Omega_b,\ \Omega_m$ to describe
the fraction of baryons and total non-relativistic matter relative to
the critical density.

\section{Estimation of gas fraction.}

We will use the observed emission weighted temperatures $<T>_X$,
central surface brightness $\Sigma_0$, angular core radii $r_c$, and
radial power law index $\beta$ to determine the desired parameters.
In order to apply the virial theorem and hydrostatic equilibrium, we
need to further assume that gas traces mass, which has been found to
be a good approximation to simulated clusters.

\begin{figure}
\plotone{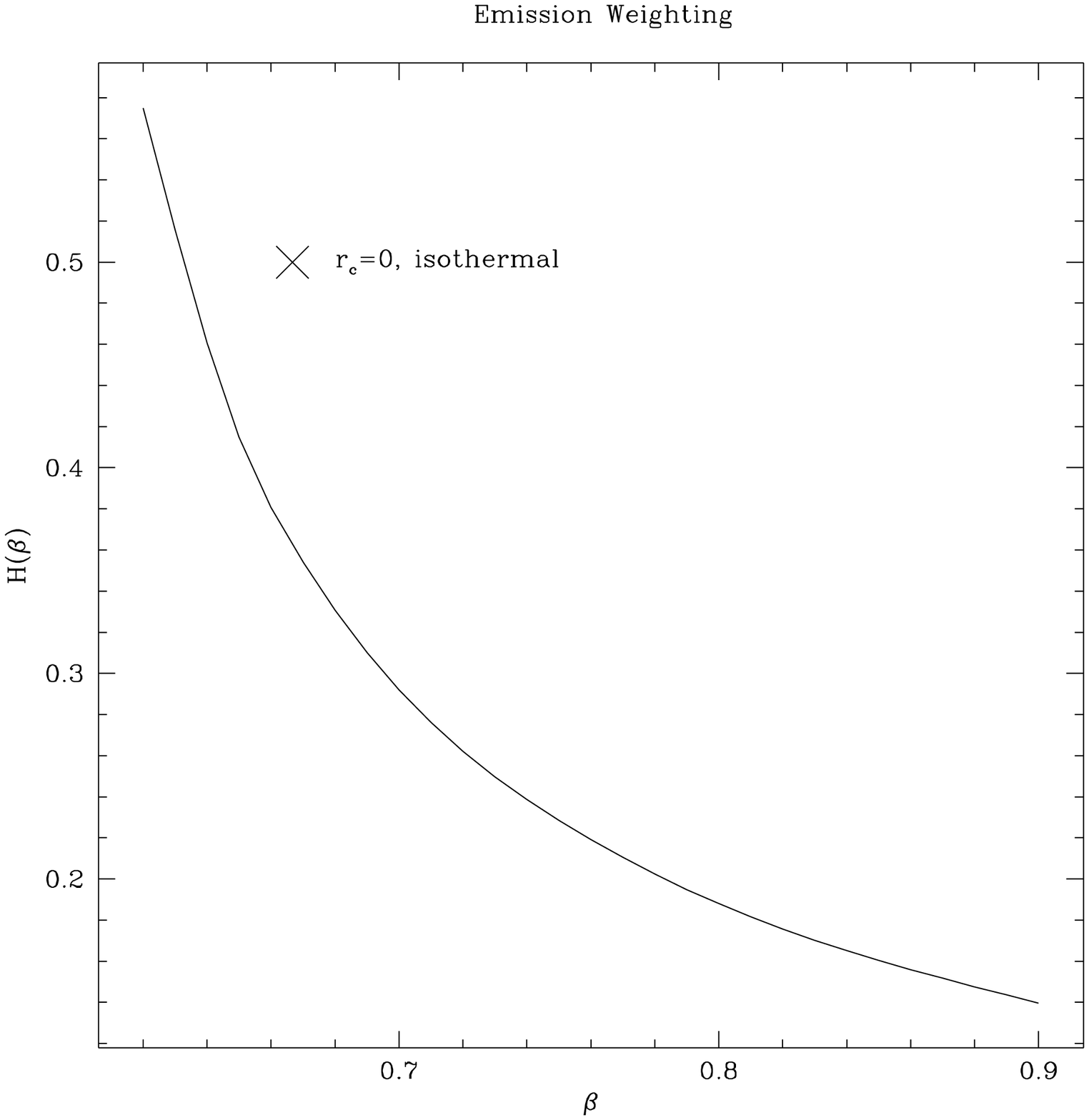}
\caption{The dimensionless weighting factor as a function of radial
profile.  The cross indicates the value of an isothermal sphere of
zero core radius $\beta=2/3, \ H(\beta=1/2)$.}
\label{fig:h}
\end{figure}

For x-ray clusters, a popular fit to the density profile is that of a
$\beta$-model (\cite{jones84}) which assumes spherical symmetry of the mass
distribution.  The model fits the density profile to be
$\rho=\rho_0(1+r^2/r_c^2)^{-3\beta/2}$.  $\rho_0$ is the central
density and $r_c$ is called the core radius.
Projected onto two dimensions,
we obtain the surface density
$\Sigma=\Sigma_0(1+r^2/r_c^2)^{-3\beta/2-1/2}$.  If we assume the
sphere to be self-gravitating, we obtain the pressure
\begin{equation}
P=4\pi G\rho_0^2r_c^2\int_{r/r_c}^\infty (1+u^2)^{-3\beta/2} u^{-2}
\int_0^u (1+v^2)^{-3\beta/2}v^2dvdu .
\end{equation}
At $r\gg r_c$, we note that $P \propto r^{-6\beta+2}$, and for typical
$\beta \approx 2/3$, the pressure drops rapidly.  Therefore the
behavior of the cluster far outside the core radius does not affect
the interior pressure significantly.  Similarly, we can derive the
temperature profile $T=\mu m_p P/\rho k$, and we can express the
central density in terms of the luminosity weighted temperature
assuming pure bremsstrahlung
$<T>_X \equiv \int \rho^2 T^{3/2}/\int \rho^2 T^{1/2}$ to be
\begin{equation}
\rho_0 = \frac{k <T>_X}{4\pi G r_c^2 \mu m_p H(\beta)}
\end{equation}
where $H(\beta)$ is shown in figure \ref{fig:h}.  We note that this
formulation is only accurate for $\beta > 0.65$, and for $\beta<4/7$
the emission weighted temperature $H(\beta)$ formally diverges.  Using
standard values for the parameters, we can now calculate the gas
fraction
\begin{equation}
f_g = 9.883 H(\beta) \left(\frac{n_e}{10^{-3} \rm cm^{-3}}\right)
\left(\frac{r_c}{\rm Mpc}\right)^2 \left(\frac{\rm keV}{T}\right)
h_{50}^{-3/2},
\label{eqn:gasfrac}
\end{equation}
where $n_e$ is the central electron density.  We have assumed the mean
molecular weight to be $\mu=0.63$.

At cosmological distances, we can similarly invert (\ref{eqn:gasfrac})
to yield the core radius.  The dimensionless deceleration parameter is
defined as $q_0=-\ddot{a} a/\dot{a}^2$ in terms of the scale factor
$a=1/(1+z)$.   In order to facilitate
translations of observed and published data, we assume that the
observational quantities were fitted using $h_{50}=1,\ q_0=0.5$ in a
matter dominated universe, as the majority of authors seem to prefer.
The observable quantities are the central surface brightness, angular
core radius and temperature.  Only the second quantity has a
cosmological $q_0$ dependence.  In order to convert the central
X-ray surface brightness into a density, we need to divide the surface
brightness by a length, and take its square root.  This introduces a
dependence on angular diameter distance to the inverse one half
power.  The $r_c$ term in  (\ref{eqn:gasfrac}) depends on the square
of the angular diameter distance.  Putting it together, we find a
dependence on the angular diameter distance to the $3/2$ power.

We have now derived the cosmological dependence
\begin{equation}
f_g(q_0)=\left(1+(\frac{1}{4} -
\frac{q_0}{2})z\right)^{3/2} \times f_g(q_0=\frac{1}{2})
\label{eqn:q0}
\end{equation}
which is independent of the Hubble constant $h$.
Strictly speaking equation (\ref{eqn:q0}) is only correct in the limit
$z\rightarrow 0$.  If the deceleration were constant $q=q_0$,
for example if $\Omega_m=1$ 
or $\Omega_m=0$, then equation (\ref{eqn:q0}) is accurate to better than
10\% for $z<4$.

\section{Sample data set.}

To implement this simple prescription, we used the Jones and Forman
cluster sample (\cite{jones84}).  In order to obtain an accurate data
set, we restricted ourselves to clusters where $L_X>10^{44}$ erg/sec,
$\beta>0.6$ and $r_c$ is determined to better than a factor of 2.
This leaves is with a set of five clusters: A85, A2199, A2255, A2256
and A2319.  The respective gas mass fractions derived from formula
(\ref{eqn:gasfrac}) are 0.22, 0.14, 0.15, 0.17, 0.22.  In addition, we
used the data from (\cite{briel92}) for the COMA cluster in
order to compare our method to the well studied numbers from White et
al (\cite{white91}).  That data yielded a gas fraction of 0.17, which is
consistent with the previous estimates by White et al.  All numbers
are expressed for $h_{50}=1$, and scale as $h_{50}^{-3/2}$.  If we use
equal weighting, the mean gas fraction is $f_g=0.178$ with a standard
deviation of $\sigma=0.034$.  That would imply a 20\% RMS cluster to
cluster variation in the gas fraction.  The mean $f_g$ estimate from
our sample of $N=6$ clusters would then be accurate to
$\sigma/\sqrt{N-1} \approx 0.015$.  This fluctuation is consistent
with the simplest hypothesis that the gas fraction is constant for all
clusters in the sample, with the variation arising solely from the
error in measurement of the cluster properties.  Detailed weighting of
the observational error bars are difficult to implement from the
published data since each of the error terms are correlated.
Nevertheless, we see that this crude method yields robust results with
a small scatter, which allows us to apply it to measure cosmological
parameters. 

The highest redshift clusters satisfying the same selection criteria
that our literature search revealed were A370 at a redshift $z=0.373$
(\cite{bautz94,struble87}) and CL0016+16 at $z=0.5545$ (Neumann and
B\"ohringer 1996).  For A370, the authors of that paper were able to
measure the temperature accurately, but due to the nature of the ASCA
X-ray satellite, the spatial profile is still subject to
uncertainties.  We use their best estimates $r_c=0.25$ Mpc, $T=8.8$
keV, $\beta=0.65$ and $n_e=6.5\times 10^{-3}$ cm$^{-3}$.  Using
(\ref{eqn:gasfrac}) and (\ref{eqn:q0}) implies
$f_g=0.189(1+(\frac{1}{4} - \frac{q_0}{2})z)^{3/2}$.  If we assume
that the gas fraction is drawn from the same distribution as the local
sample, we obtain $q_0 = 0.71^{+0.68}_{-.64}$ with $1-\sigma$ errors.

Cl0016+16 is a better candidate due to its higher redshift.  Neumann
and B\"ohringer used both the ROSAY HRI and PSPC to measure the
cluster parameters.  For the HRI the best fit values are $\beta=0.8,\
r_c=.372$Mpc, $n_e=6.5\times 10^{-3}$cm$^{-3}$, from which we deduce a
gas fraction $f_g=.20$ for $q=1/2$.  The PSPC parameters are slightly
different, $\beta=.68,\ r_c=.283$Mpc, $n_e=7.6\times
10^{-3}$cm$^{-3}$, for which $f_g=.24$.  We use the average of those
two values $f_g=.22$, and apply the same procedure as above to obtain
$q_0=.98^{+.41}_{-.39}$, again at $1-\sigma$.  The HRI and PSPC gas
fractions vary from their average by 10\%, which is smaller than the
local scatter of 19\% which we used to determine the error bars.  It
is difficult to quantify how errors should be added or treated, and
they should not be considered rigorous in any sense.  While the
published values for the $\beta$ model parameters include error bars,
these are strongly correlated.  The model depends only strongly on the
asymptotic radial profile of the gas, which is well determined in the
fits.  The observational errors in the determination of $r_c$ and
$n_e$ are correlated in such a way as to preserve the asymptotic
radial behavior, $n_e \sim r_c^{3\beta}$.  In principle it is possible
to obtain intrinsic errors by fitting the models directly to the raw
data. For now, the most robust estimator is still the local sample.
Some systematic errors could of course enter, the largest of which
being an inherent evolution in the gas fraction of clusters.
Neglecting such possibilities allows us to formally combine the two
observations and derive a combined $q_0=.91\pm .34$

A more difficult example is RXJ1347.5-1145 (\cite{schind96}).  At a
redshift $z=0.451$ this cluster was found to have a strong cooling
flow, and formally $\beta=0.56$.  Our formula (\ref{eqn:gasfrac})
diverges, and therefore does not apply for this value of $\beta$.
While a significant portion of the emission may come from the cooling
flow (the authors estimate 43\%), the total gas mass interior to $240$
kpc is well constrained.  The authors obtained a gas fraction at that
radius for a flat universe of $f_g=0.19$.  Their inferred gas mass
fraction increases with radius since the dark matter was assumed to be
isothermal, while the gas was assumed scale as $\rho \propto
r^{-1.68}$.  Due to a cooling flow at the center of the cluster, the
radial profile cannot be inferred accurately.  The gas mass is only
directly well determined at the 240 kpc radius, even at 1 Mpc the
uncertainties are large, probably more than a factor of two.  Our {\it
Ansatz} assumed that the gas fraction is independent of radius, so we
would infer here at the most accurate radius that $f_g(q_0=0.5)=.19
h_{50}^{-3/2}$.  Applying (\ref{eqn:q0}) we find
$q_0=.69_{-.52}^{.56}$ for this cluster.

Combining all three clusters now formally gives $q_0=.85 \pm .29$.
Conversely, if we allow the cluster gas fraction to vary freely in
time and assume, then a $q_0=-1/2$ would imply that clusters of
galaxies at a redshift $z\sim.5$ had a gas fraction 50\% higher than
clusters do today.  We note that the distant cluster sample only had a
small scatter in their relative gas fraction, no bigger than that of
the local sample.

\section{Systematic Uncertainties.}

Historically the measurement of $q_0$ has been plagued by evolutionary
effects (\cite{mtw}).  Objects at cosmological distances represent the
universe when it was younger, and could certainly have been
systematically different in the past.  It has been difficult to
measure the evolution of objects separately from the evolution of
space-time.  The current estimate relies on the gas fraction  remaining
constant, which is justified by the similarity in the equation of
state between collisionless dark matter and an ideal gas, as well as
numerical simulations which confirm these guesses.  But here we do not
have to believe blindly.  Near term high resolution temperature
measurements with the AXAF X-ray satellite to be launched in 1998 will
measure the gas fraction of the local population to high accuracy.
Recent measurements of gravitational lensing have been able to
reconstruct the mass distribution in distant clusters, in fact a
redshift of 1/2 is ideal for such measurements.  The strong arcs
observed in A370 (\cite{gro89}) would certainly be compatible with an
assumed constant gas fraction, but here we still have systematic
uncertainties involved.  Comprehensive modelling combining weak and
strong lensing can ultimately test the assumption of constant gas
fraction between clusters at similar distance, and constancy in the
radial profile.

The cluster sample was chosen from the literature, which does not have
an objective selection function.  If some clusters had a higher gas
fraction at some fixed temperature, these would also be most X-ray
luminous and might be preferably selected.  It is not clear that the
distant cluster sample had similar selection criteria as the local
sample.  It is therefore by no means statistically fair to simply
compare the averages of gas fractions in these two populations and
draw definitive conclusions.  In the absence of a quantitative
selection function, however, we can only ignore the possibility of
this effect.  The only reassurance is that the scatter in cluster gas
fractions is small, both in the local as well as in the distant
sample.

We have assumed that the gas is spherically symmetric and in
hydrostatic equilibrium.  Since the cores of clusters are overdense by
about $10^4$ relative to the cosmic density, the age of the cluster is
a hundred fold longer than the dynamical time $t_d\approx
1/\sqrt{G\rho}$.  Since any perturbations are expected to damp on a
dynamical time, one would expect the assumptions of hydrostatic
equilibrium to hold to high accuracy.  Numerical
simulations (\cite{pen96}) indicate that the kinetic energy fraction in
the core is indeed typically less than 10\% of the thermal energy.  We
can make some estimates to the expected presence of substructure.
\cite{henry92} observed no evolution in the cluster
luminosity function to a redshift of $z=0.35$.  We can thus assume
that the clusters are at least half as old as the universe.  Let us
consider two extreme cases.  If they grew by a constant rate of
continuous accretion, we would find all clusters with the same amount
of non-equilibrium energy of a few percent.  The opposite extreme is
that clusters formed by sudden mergers of equal mass objects.  In that
case, a few percent of clusters would be just merging, and be
maximally out of equilibrium.  The remaining ninety-some percent would
be in very good hydrostatic equilibrium.  The truth is probably
somewhere inbetween.  There may be a few clusters just undergoing
mergers, and all clusters may have some small amount of substructure.
But we see that the effect must be quite small, unless we are unlucky
and happen to be selecting the few non-relaxed clusters.  We would
need to wait until we collect a larger sample size at cosmological
redshift before we can know for sure.  Occasionally subclumps could
appear near the cores of clusters due to chance projections.

Similarly, one would expect the core to be more spherical than the
cluster as a whole, since the core has had more time to undergo
relaxation.  The assumptions of hydrostatic equilibrium can be tested
using the X-ray satellite AXAF by comparing line ratios to line
widths.

Core radii are often difficult to measure accurately, and the formal
error bars on the fits in the local sample is larger than the inferred
gas fraction scatter.  The reason for this behavior is an almost
perfect correlation between the error in fitting the core radius and
the inferred central density.  The parameter which is usually well
determined is the surface brightness at radii slightly larger than the
core radius, where most photons are measured.  Errors in the core
radius and central density will tend to correlate in a way to preserve
this surface density.  In the case that $\beta=2/3$, which is in the
typical range for clusters, any such correlated change have no effect
at all on the total gas fraction.  Of course, the $\beta$ model
parametrization provides a convenient notation to describe the density
outside the core radius, but if we wish to have uncorrelated
measurements, one should perhaps parametrize the surface density just
outside the core radius separately.

These arguments might explain the small observed scatter in the gas
fraction.  There are undoubtably an even greater number of effects
which would contribute to increase the measured scatter, but
empirically we have found that all those effects either cancel, or
contribute at most 20\% scatter to the deduced gas fraction.

In simulations (\cite{pen96}) we find that the dark matter is more
centrally concentrated than the gas, even though they trace each other
quite well beyond a core radius.  This is also found in lensing
studies (\cite{kneib96}), and can be explained in terms of the second law
of thermodynamics for gases.  While the entropy of a fluid element
must increase along its flow line (in the absence of cooling), the
dark matter has no such constraint.  On the contrary, the latter
settles to high density cores due to the process of violent
relaxation, which locally violates the second law of thermodynamics.
During the formation of a cluster in a time dependent potential, the
fast moving particles leave the center, while the slow ones
preferentially remain there.  This Maxwellian demon leads to more
centrally concentrated dark matter as the cluster forms.  The sign of
this effect is to lower the velocity dispersion of the dark matter
relative to the gas interior to the core radius, which means we might
systematically underestimate the gas fraction using the procedure
described here.  In simulations, this effect would affect our estimate
by $\approx$ 20\%, but this offset is independent of redshift.  Once
high resolution resolved temperature profiles are measured with AXAF,
we can test the structure of actual clusters.

At radii much smaller than the core radius it appears that gas does
not trace total mass.  While clusters are well fit by the $\beta$
model with well defined core radius $r_c \gtrsim 200 h_{50}^{-1}$ kpc
(Jones and Forman 1984), lensing indicates that the dark matter
continues to increase in density inward (Kneib \etal 1996).  This had
led to some activity, including questioning of the hypothesis of
hydrostatic equilibrium (Babul and Katz 1993, Loeb and Mao 1994).  The
presence of lenses were inconsistent with an isothermal gas in
hydrostatic equilibrium at the measured temperature.  As Babul and
Katz point out, the discrepancy can be resolved completely if one
allows for an 80\% increase in the temperature from the core radius at
$r_c = 230 h_{50}^{-1}$kpc to the arc radius $\theta_a =
80h_{50}^{-1}$kpc.  When emission weighted, the increase in the
projected temperature profile is only 49\%, where $\sim 5\%$ of the
photons would originate from the hotter region, and may not be
currently detectable.  By itself, this would imply an entropy
inversion, and such a cluster would be convectively unstable.
However, a combination of temperature gradient and density gradient
within the observational errors could reduce or eliminate the
discrepancy.  Additional masses behind the cluster could also
contribute to the inferred gravitational mass.  Moreover, weak lensing
studies for A2218 (Squires \etal 1996a) yield mass estimates
consistent with the gas mass estimates at the arc radii.

The method proposed in this article depends on observable quantities
near the core radius.  This makes the estimate robust, and less
dependent on cosmological uncertainties.  At larger radii, the
assumption of hydrostatic equilibrium and spherical symmetry breaks
down, and the clusters are observed to have significant substructure.
At smaller radii, the discrepancy between dark matter and gas
increases, as we discussed above.  Clusters may provide a handle as a
cosmic meter in the region near the core radius.

There is an additional possibility that the gas fraction in clusters
changed due to non-gravitational processes, such as heating, cooling,
or gas injected from galaxies.  While cooling flows are an interesting
phenomenon observed in several clusters, most clusters do not exhibit
cooling flows.  Current estimates indicate that up to 30-50\% of X-ray
clusters may have cooling flows, amongst which 10\% of the X-ray
emission may arise from such an effect (Fabian 1994).  These effects
are small compared to the error budget in our estimates.  They would
lead to a slight increase of the gas fraction at the center of the
cluster core, violating the assumption stated earlier that gas traces
total mass at all radii.  Gas injection from galaxies would cause the
gas fraction to increase in time, raising the inferred value of $q_0$.
Heating could lower the gas fraction in clusters today.  The primary
signature of such an effect would be that colder clusters should have
a much lower gas fraction than hot clusters.  In fact, the local
sample shows no dependence of gas fraction on cluster temperature.

Squires \etal (1996b) used weak lensing to measure the radial mass
profile for A2163 at $z=.201$, and found it consistent with the gas
tracing total mass at all radii.  They inferred the mean gas fraction
to be $f_g=.2 h_{50}^{-3/2}$, consistent with the local sample.  At
this redshift, no constraints on $q_0$ are obtained with this cluster.
Similarly, Squires \etal (1996a) found that gas traces total mass for
A2218 at $z=.17$, with $f_g \sim .11 \pm .57 h_{50}^{-3/2}$, still
consistent with our hypothesis.  While weak lensing probes the mass
profile well, it allows a systematic error in normalizing the absolute
mass.  Thus the current data appears consistent with all our assumptions.

\section{Conclusions}

We have presented a new promising method for measuring angular diameter
distance and determining the deceleration parameter $q_0$.  Using the
{\it Ansatz} of constant gas fraction, we are able to obtain a small
scatter of only 20\% in the observed local cluster sample selected
objectively from a published catalog (Jones and Forman 1984).  This is
sufficient to obtain interesting cosmological constraints using only
three clusters at moderate redshift $z\sim 0.5$.

Using data which is currently being collected by ASCA and ROSAT, the
cluster sample will increase tenfold.  Most importantly for the
future, we need a systematic study of cluster gas fraction from
samples with clear selection criteria, such as an X-ray flux limited
sample.  Weak lensing data is improving rapidly for clusters at
redshifts near z=0.4, which in the near future may provide us with an
accurate measure of the deceleration parameter $q_0$ (Tyson and
Fischer 1995, Squires \etal 1996a,b).  In the longer term future when
AXAF is launched, more detailed modeling will become possible.  The
current analysis is still subject to large and systematic
uncertainties, but we have demonstrated that even with the limited
available published data, interesting cosmological constraints can be
derived.  This strategy will provide an independent measure of $q_0$
to compare with supernovae searches (\cite{perl96}) and the
Sunyaev-Zeldovich effect (\cite{whi78}).

If gas traces mass, as simulations suggest, then $\Omega_b/\Omega_m >
f_g$.  The high gas gas fraction $f_g \approx 0.178 h_{50}^{-3/2}$
together with big bang nucleosynthesis $\Omega_b h_{50}^2 = 0.05$
(\cite{walker1991}) would imply a dark matter fraction $\Omega_m <
0.28/\sqrt{h_{50}}$.  Such a low abundance of dark matter would
suggest the presence of possibly a cosmological constant or negative
spatial curvature.  The curvature scenario predicts a deceleration
parameter $0< q_0 < 1/2$, while the cosmological constant would imply
$q_0 \la -1/2$.  A low value of $\Omega_0$ has also been proposed for
several other reasons (\cite{ost95}), but the choice between a
cosmological constant or spatial curvature has so far been
aesthetical.  A measurement of $q_0$ would ultimately distinguish
these scenarios.  Using the small sample described above the
hyperbolic spatial curvature model has a higher probability of fitting
the observations, but we cannot rule out a cosmological constant with
the available data.

{\sc Acknowledgements}.  I thank Avi Loeb, Tsafrir Kolatt, Alexey
Vilkhein, Christine Jones and William Forman for useful discussions.
This work was supported by the Harvard Society of Fellows and the
Harvard-Smithonian Center for Astrophysics.

\newpage

\end{document}